\documentclass[preprint2]{aastex}

\def\gtsim{\mathrel{\hbox{\rlap{\hbox{\lower4pt\hbox{$\sim$}}}\hbox{$>$}}}}
\def\lesssim{\mathrel{\hbox{\rlap{\hbox{\lower4pt\hbox{$\sim$}}}\hbox{$<$}}}}

\shorttitle{Kinematics of late stars in the Galactic disk}
\shortauthors{B. Fuchs et al.}

\begin{document}


\title{The kinematics of late type stars in the solar cylinder \\ 
       studied with SDSS data}


\author{Burkhard Fuchs\altaffilmark{\ref{ARI}},
Christian Dettbarn\altaffilmark{\ref{ARI}},
Hans-Walter Rix\altaffilmark{\ref{MPIA}},
Timothy C. Beers\altaffilmark{\ref{MSU}},
Dmitry Bizyaev\altaffilmark{\ref{APO}},
Howard Brewington\altaffilmark{\ref{APO}},
Hartmut Jahrei{\ss}\altaffilmark{\ref{ARI}},
Rainer Klement\altaffilmark{\ref{MPIA}},
Elena Malanushenko\altaffilmark{\ref{APO}}, 
Viktor Malanushenko\altaffilmark{\ref{APO}},
Dan Oravetz\altaffilmark{\ref{APO}},   
Kaike Pan\altaffilmark{\ref{APO}}, 
Audrey Simmons\altaffilmark{\ref{APO}},
Stephanie Snedden\altaffilmark{\ref{APO}}
}

\altaffiltext{1}{Astronomisches Rechen-Institut am Zentrum f\"ur Astronomie der
Universit\"at Heidelberg, 69120 Heidelberg, Germany \label{ARI}}
\altaffiltext{2}{Max-Planck-Institut f\"ur Astronomie, 69117 Heidelberg,
Germany \label{MPIA}}
\altaffiltext{3}{Department of Physics and Astronomy, CSCE: Center for the 
Study of Cosmic Evolution, and JINA:  Joint Institute for Nuclear Astrophysics,
Michigan State University, E. Lansing, MI 48824, USA\label{MSU}}
\altaffiltext{4}{ Apache Point Observatory, Sunspot, NM, 88349,
USA\label{APO}}


\begin{abstract}
We study the velocity distribution of Milky Way disk stars in a
kiloparsec-sized region around the Sun, based on $\sim 2$
million M-type stars from DR7 of SDSS, which have
newly re-calibrated absolute proper motions from combining
SDSS positions with the USNO-B catalogue. We estimate
photometric distances to all stars, accurate to $\sim 20\%$,
and combine them with the proper motions to derive
tangential velocities for this kinematically unbiased
sample of stars. Based on a statistical de-projection method
we then derive the vertical profiles (to heights of $Z$ = 800 pc above
the disk plane) for the first and second moments of the three
dimensional stellar velocity distribution. We find that $\langle W \rangle$ =
$-7$ $\pm$ 1 km/s and $\langle U \rangle$ = $-9$ $\pm$ 1 km/s, independent
of height above the mid-plane, reflecting the Sun's motion with
respect to the local standard of rest. In contrast, $\langle V \rangle$ changes
distinctly from $-20$ $\pm$ 2 km/s in the mid-plane
to $\langle V \rangle$ = $-32$ km/s at $Z$ = 800 pc, reflecting an asymmetric
drift of the stellar mean velocity that increases with height.
All three components of the M-star velocity dispersion show a
strong linear rise away from the mid-plane, most notably
$\sigma_{ZZ}$, which grows from 18 km/s ($Z$ = 0) to 40 km/s (at $Z$ = 800 pc).
We determine the orientation of the velocity ellipsoid, and
find a significant vertex deviation of 20 to 25 degrees, which
decreases only slightly to heights of $Z$ = 800 pc.
Away from the mid-plane, our sample exhibits a remarkably large
tilt of the velocity ellipsoid towards the Galactic plane,
which reaches 20$^\circ$ at $Z$ = 800 pc and which is not easily explained.
Finally, we determine the ratio $\sigma^2_{\phi\phi}/\sigma^2_{RR}$
near the mid-plane, which in the epicyclic approximation
implies an almost perfectly flat rotation curve at the
Solar radius.

\end{abstract}




\section{Introduction} \label{sec1}

The velocity distribution of stars in the Milky Way has been most
comprehensively studied in the solar vicinity. One of the most long standing
systematic collection of relevant data is the {\em Catalogue of Nearby Stars}
\citep{Glie57}, with a fourth edition ({\em CNS4}) now nearly completed 
(Jahrei{\ss}, in preparation). The {\em CNS4} contains space velocity data for a
few thousand stars. A much larger data set, based on stars with proper motions
measured by the {\em Hipparcos} satellite (ESA 1997), was analyzed by 
\cite{DeBi98}. From the proper motions of $\sim$ 10$^4$ stars they could 
reconstruct the first and second moments of the velocity distribution using a
statistical de-projection method which makes use of the many viewing directions
through the velocity distribution of the stars. The velocity dispersions of
stars in the solar vicinity were also studied by \cite{Nord04} from the {\em
Geneva-Copenhagen Survey} of F and G stars. These authors complemented the
parallaxes and proper motions of the stars in the {\em Hipparcos} catalogue by
their own radial velocity measurements. With the advent of the deep, wide-field
{\em SDSS} and {\em RAVE} surveys it has become possible to study the 
kinematics of stars in the Galactic disk at much larger distances from the Sun.
Recently \cite{Boc07} have drawn from the {\em Fifth Data Release of the Sloan
Digitized Sky Survey} \citep{Adel07} a sample of about 7000 K and M stars for
which three-dimensional space velocities are available (the SLoMaSS sample). 
These stars were selected
along a line-of-sight which reaches a distance of 1 kpc above the Galactic
midplane. \cite{Velt08} and \cite{KlFu08} have analyzed the {\em First Data 
Release of the RAVE Survey} \citep{Stei06}. \cite{Velt08} inferred the 
kinematics of stars in the solar cylinder by modeling the distribution of stars 
in phase space, which was then constrained by star counts of the {\em RAVE} 
data. \cite{KlFu08} searched for fine structure in local phase space. 
Most recently, \cite{Sie08} have analyzed the
velocity distribution of red clump giants at heights between 500 and 1500 pc
above the Galactic midplane, using data from the {\em Second Data Release of the
RAVE Survey} \citep{Zwi08}.

In this paper we analyze stellar data from the {\em Seventh Data Release of the
Sloan Digital Sky Survey (SDSS)} \citep{AAA08}, which includes imaging taken as
part of the {\em Sloan Extension for Galactic Understanding and Exploration
(SEGUE)} program \citep{Yan08}, one of three sub-surveys that were part of
first extension of the {\em SDSS}, known as {\em SDSS-II}. The {\em SEGUE}
program was designed, in part, to obtain $ugriz$ imaging of some 3500 square 
degrees of sky outside of the original {\em SDSS-I} footprint 
(\citealt{Fuk96}, Gunn et al.~1998, 2006, \citealt{Yor00}, \citealt{Hog01},
\citealt{Smi02}, \citealt{Sto02}, Abazajian et al.~2003, 2004, 2005,
\citealt{Pie03}, \citealt{Ive04}, Adelman-McCarthy et al.~2006, 2007, 2008,
\citealt{Tuck06}. The regions of
sky targeted are primarily at lower Galactic latitudes 
($|b| \lesssim 35^{\circ}$), in order to better sample the disk/halo interface
of the Milky Way. {\em SEGUE} also
obtained $R$ $\simeq$ 2000 spectroscopy, over the wavelength range $3800-9200$\,
{\AA}, for some 250,000 stars in 200 selected areas over the sky available from
Apache Point, New Mexico \citep{Yan08}. The spectroscopic candidates are
selected on the basis of $ugriz$ photometry to populate target categories
chosen to explore the nature of the Galactic stellar populations at distances
from 0.5 kpc to over 100 kpc from the Sun. The {\em SEGUE Stellar Parameter 
Pipeline} ({\em SSPP}; see Lee et al.~2008a,b, \citealt{All08}) obtained 
estimates of radial velocities, and in particular, the stellar atmospheric
parameters, for the targeted stars over wide ranges of effective temperature
($T_{\rm eff}$), surface gravity (log $g$), and metallicity ([Fe/H]).

Of central importance to the present paper, the {\em DR-7} catalog contains
astrometric and photometric ($ugriz$) data for well over 100 million stars. Many
of these have been identified in the {\em USNO-B} catalog of proper motions
\citep{Mone03}. These measurements have been improved through a re-calibration
effort based on {\em SDSS} data, and provide us with an extensive set of
kinematical data \citep{Munn04}. After correction of a small systematic error in
the data reduction procedure (Munn 2007, internal {\em SDSS} memorandum;
note that the {\em DR-7 CAS} proper motions have been corrected for this error),
the absolute
accuracy of the proper motions has reached now a few milliarcseconds per year,
corresponding to, e.g., $\sim$5 km/s at a distance of 530 pc for $\epsilon_\mu$
= 2 mas/year. They are thus superbly suited to study the kinematics of the
Galaxy, especially for stars that populate a local volume.

We have extracted information for about 2 million M-type stars in the solar
cylinder from the {\em SDSS DR-7} database; these stars represent a fair sample
of an old, relaxed stellar population mostly drawn from the Galactic thin disk.
Using a photometric distance estimator we convert the proper motions into
tangential velocities. For individual stars, tangential velocities alone do not
allow the construction of space velocity components because the radial
velocities are missing. However, \cite{DeBi98} have shown that, by making use of
the many viewing angles through the velocity distribution of stars over a large
fraction of the sky, the characteristic moments of the velocity distribution can
be determined in a statistical manner. Following this method, we are able to
determine empirically, {\rm for the first time with high precision}, the first
and second moments of the velocity distribution of local stars as a function of
height above the Galactic midplane. 

Our paper is organized as follows. In \S 2 we describe the sample selection and
our methods for obtaining distances and deriving tangential velocities. In \S 3
we apply the statistical approach of \cite{DeBi98} to reconstruct the moments of
the three-dimensional velocity distribution of stars from their tangential
velocities. Our conclusions are summarized in \S 4. 


\section{Sample Selection, Distance Estimates, and Tangential Velocities} 
\label{sec2}

In order to obtain a sample of stars that are representative of the old, 
relaxed population dominated by the thin disk of the Milky Way, we have 
extracted from the 
{\em SDSS DR-7} database all objects classified as stars in the {\em
de-reddened} color range $0.49\,\leq\, r-i\,\leq \,1.6$, $1.25 \, 
< \, g-r \, < 1.50$, and with apparent magnitudes 
$15 \, < g < \, 20.5$ (cf.~\citealt{IvBo05}).
These stars have spectral types M0 and later. The upper cutoff at $g$ = 20.5
mag, which corresponds to $r$ = 19.5 mag, is necessary because the proper
motions are seriously contaminated by misidentifications at fainter magnitudes
(Lepine 2008, internal SDSS memorandum). We have selected only stars with
non-zero proper motion errors, apparent magnitude errors $\leq$ 0.05 mag in the
$g,r,i$ bands, and which were not flagged for any known astrometric
or photometric peculiarities. This resulted in a sample of about 2.6 million
stars (mean age expected to be $\sim$5 Gyr).

\begin{figure}
\plotone{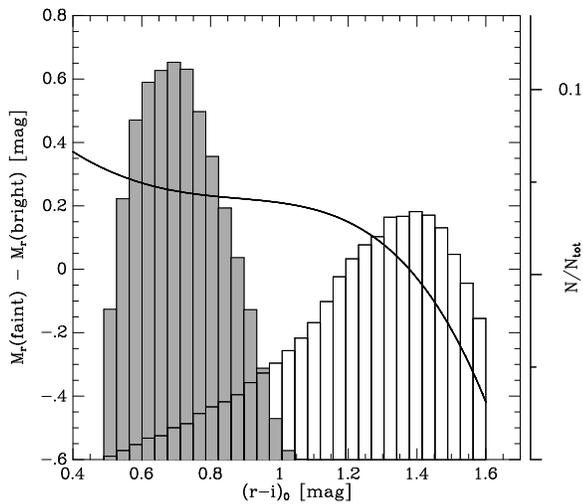}
\caption{Difference between the faint and the bright normalization of the 
absolute magnitude $M_r$-color relation by \cite{Juri08}, as a function
of $(r-i)_0$ color (solid line). Underlaid as white histogram
is the frequency distribution of the M stars in the distance bin closest to
the Galactic midplane, while the grey histogram shows the corresponding 
distribution in the most distant bin. Both histograms are normalized to unity
(scale on the right).}
\label{fig-1}
\end{figure}

We have determined a photometric distance estimate for each star using the 
absolute magnitude-color relations of \cite{Juri08}. We have compared both
the `faint' and `bright' normalizations with color-magnitude relations by
\cite{Dave06} and \cite{West05}. \cite{Dave06} have observed M stars
drawn from the {\em Catalogue of Nearby Stars} \citep{JaWi97}, for which 
trigonometric parallaxes are available, in the $ugriz$ filter system. 
\cite{West05} have selected M dwarfs from the {\em Third Data Release of the
Sloan Digital Sky Survey} \citep{Abaz05}
and determined spectroscopic parallaxes of these stars. If we
use fits to the \cite{West05} or \cite{Dave06} data as distance estimators in
the de-projection formalism (described below) to obtain the first and second
moments of the velocity distribution of the stars, we find that the results are
practically indistinguishable from the results based on the \cite{Juri08} bright
normalization. We conclude that this normalization is
thus well suited to serve as a photometric distance estimator for red main
sequence stars in the Galactic thin disk, such as M dwarfs,
\begin{eqnarray}
M_r &= &3.2 +13.30\,(r-i)-11.50 \,(r-i)^2 \nonumber \\ 
 & & +5.40\,(r-i)^3 -0.70\,(r-i)^4 \,.
\label{eq1}
\end{eqnarray}
As explained below, we count M dwarfs up to distances reaching 800 pc above the 
Galactic midplane. According to the model of the Galactic thin and thick disks 
by \cite{Juri08}, we expect about 13\% of the M dwarfs close to the midplane to 
be thick-disk stars; in the distance range of 600 to 800 pc above the
midplane the fraction of thick disk stars, which are more metal-poor than thin
disk stars, is expected to be about 36\%. Therefore,
we have compared the faint and bright
normalizations of \cite{Juri08} with the fiducial line delineating the main
sequence of the globular cluster M71 as observed by \cite{Clem08}, 
transformed into the native $ugriz$ system using the relations of \cite{Tuck06},
and assuming a distance modulus of $(m-M)_0$ = 13.02. M71 has a metallicity of
$[Fe/H]$ = -0.73 \citep{Har96} which is typical for the majority of
thick-disk stars. This
comparison shows that for metal-poor stars in the color range $r-i$ $>$ 0.5,
photometric distances can be reliably determined using the faint
normalization,
\begin{eqnarray}
M_r &= &4.0 +11.86\,(r-i)-10.74 \,(r-i)^2 \nonumber \\ 
 & & +5.99\,(r-i)^3 -1.20\,(r-i)^4 \,.
\label{eq1a}
\end{eqnarray} 
In the following we provide results using both distance scales in order to
demonstrate how (little) our results are systematically affected by the
presence of thick-disk stars. The difference between the two calibrations is 
illustrated in Fig.~\ref{fig-1} as function of de-reddened $(r-i)_0$ color.

The right ascension and 
declination, taken together with the distance estimate, were used to calculate 
the position of each star in a Cartesian reference system centered on 
the position of the Sun, with the $X$-axis pointing toward the Galactic center, 
the $Y$-axis in the direction of Galactic rotation, and the $Z$-axis towards 
the Galactic north pole, respectively. We then defined eight pill-box 
shaped counting volumes, oriented parallel to the Galactic 
midplane and stacked onto each other. Each volume has a vertical 
height of 100 pc and a radial size $\sqrt{X^2+Y^2}\leq\,1$ kpc. The midpoints 
of the counting volumes are centered on the Galactocentric radius of the Sun
and $Z$ = 50, 150, ... 750 pc, respectively. In Fig.~\ref{fig-1} we have
overlaid frequency distributions of the colors of stars in the distance bin
closest to the Galactic midplane and in the most distant bin, respectively, on
the absolute magnitude differences between the faint and bright calibrations.
These indicate which color ranges of the calibration of the absolute magnitudes
are relevant in the various distance bins.

\begin{figure}
\plotone{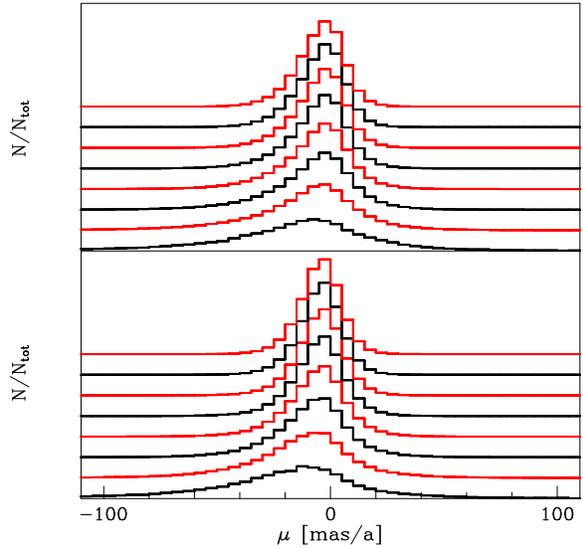}
\caption{Frequency distributions of the proper motions in right ascension (top
panel) and declination (bottom panel) for the stars in each of the eight
counting volumes. The histograms are each normalized to unity and shifted
relative to each other so that the $Z$-slice closest to the Galactic midplane
appears as lowest and the most distant $Z$-slice at the top.}
\label{fig0c}
\end{figure}

Fig.~\ref{fig0c} shows the frequency distributions of the proper motions in
right ascension and declination for the stars in each counting volume. Averaged
over the entire sample the errors of the proper motions are
$\epsilon_{\mu_\alpha}$ and $\epsilon_{\mu_\delta}$ = 3.36 mas/year.

\begin{figure}
\plotone{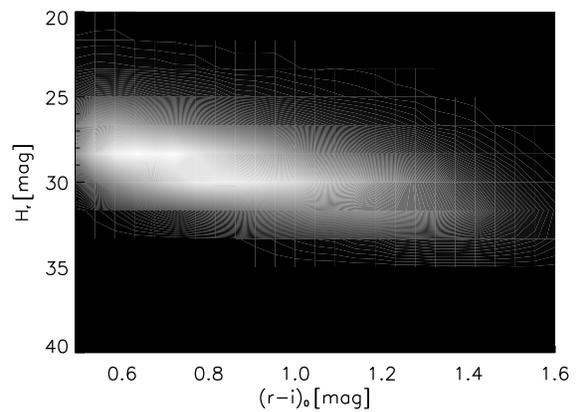}
\caption{Greyscale diagram of the frequency distribution of the reduced proper
 motions of the selected M stars versus their $(r-i)_0$ colors. The 
 distribution is shown on a linear scale, coded from black to white.}
\label{fig0}
\end{figure}

Fig.~\ref{fig0} illustrates our data as a Hess diagram  
of the reduced proper motions for our stars
\begin{equation}
H_r=r + 5\,\log{\mu} +5
\label{eq1b}
\end{equation}
versus their $(r-i)_0$ colors.
In eq.~\ref{eq1b} $\mu$ denotes the total proper motion of each star in units of
mas/year. Even though the stars have been selected into the counting volumes 
according to their estimated distances, their reduced
proper motions are to first order distance independent \citep{Luyt39}, and are 
thus also independent from the chosen distance scale. Since M giants are
intrinsically brighter (by at least 9 magnitudes) than M dwarfs, they would 
show up in Fig.~\ref{fig0} in the lower flank of the distribution. As can be
seen from Fig.~\ref{fig0}, the contamination of our sample by such halo M 
giants is, as expected \citep{Cov08}, negligible.

\begin{figure}
\plotone{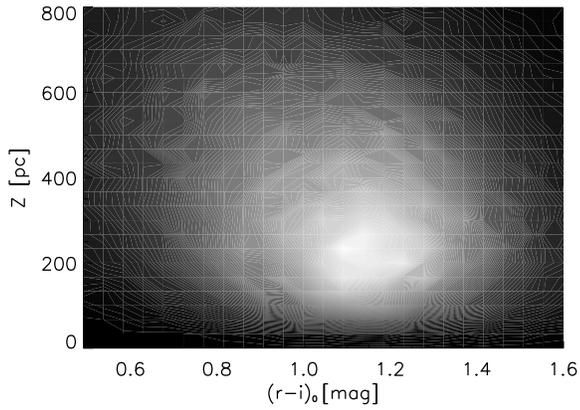}
\caption{Greyscale illustration of the distribution of the M stars in our 
sample as function of height above the Galactic midplane and $(r-i)_0$ color.
The density is shown on a linear scale, coded from black to white.}
\label{fig0b}
\end{figure}

Fig.~\ref{fig0b} shows the distribution of the selected stars as
function of height above the Galactic midplane versus their $(r-i)_0$ colors. 
As can be seen in this figure, our sample is volume limited for stars
bluer than $(r-i)_0 \leq 0.95$ (with the exception of perhaps a few bright, very
nearby stars). In the upper right portion of Fig.~\ref{fig0b},
red M stars are missing whose apparent magnitudes are fainter than the 
magnitude limit of the survey.

Returning to Fig.~\ref{fig0}, we note a mild Malmquist bias in our sample. The
high altitude counting volumes are populated by comparatively more blue stars
than red stars, and, given the apparent magnitude limits of the survey, the
blue stars are on the average intrinsically brighter.

\begin{figure}
\plotone{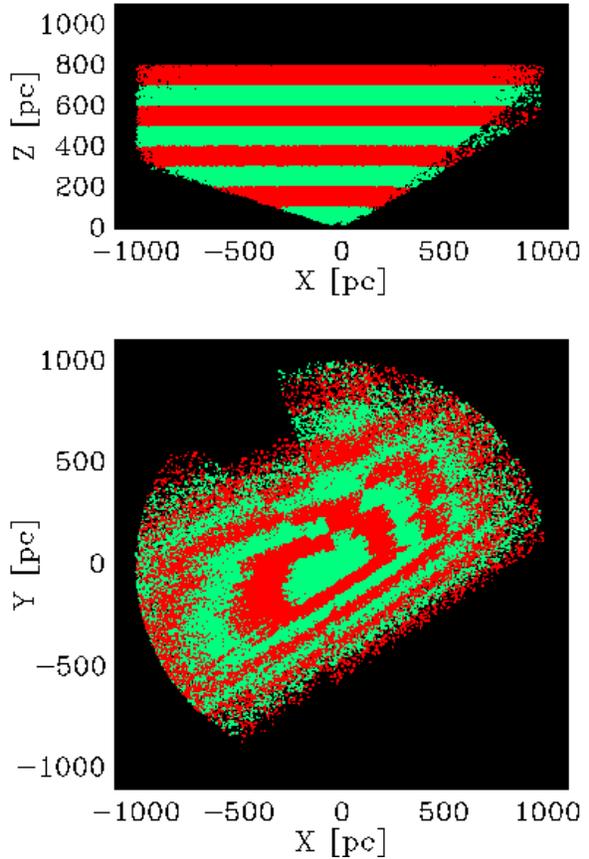}
\caption{Spatial distribution of the sample stars. The upper panel shows the
distribution in Galactocentric radius $X$ and height above the Galactic midplane
$Z$. The conical distribution reflects the geometry of the survey. The lower
panel illustrates the projection onto the Galactic plane. The layers are shown
stacked opaquely onto each other viewed from below the Galactic midplane.}
\label{fig1}
\end{figure}

For each star we have calculated its tangential velocity from the proper 
motions and the photometric distances based on the bright normalization by 
\cite{Juri08}. Assuming zero line-of-sight velocities, we then computed three-
dimensional space velocities. Following the notation of \cite{DeBi98}, these
are referred to as the proper motion velocities $p_U$, $p_V$, and $p_W$. 
The velocity components are oriented in the same way as the spatial 
coordinates. Finally, we rejected stars which had at least one single velocity 
component $>$ 150 km/s. After application of this cut, our final sample 
comprises $1\,898\,598$ stars. Their spatial distribution is shown in 
Fig.~\ref{fig1}. 
The errors of the space velocity components were determined by application of
a standard Gaussian error propagation formalism. Following \cite{Juri08},
we have assumed individual relative distance errors of 20\%, which corresponds
to an error in the absolute magnitudes of 0.43 mag. These errors result
in typical errors for the individual velocities of $\epsilon_{p_U}$, 
$\epsilon_{p_V}$, and 
$\epsilon_{p_W}$ = 2.5(+2, -1) km/s, 2.5(+3, -1) km/s, and 1.5(+3,-1) km/s
respectively, but with tails out to $\sim$ 10 km/s. However, the individual 
errors are greatly diminished when one considers the large number of stars in
each of our counting volumes. For instance, if we were to
change the individual distance errors from 20 percent to 30 percent, the 
resulting first and second moments of the velocity distribution would be 
precisely the same as given in Table 1. 

The last step in the preparation of the data was the correction of each
individual proper motion velocity for Galactic mean rotation
\begin{eqnarray}
p_U&=&p_U-(A-B)\,d \cos{(b)}\sin{(l)} \nonumber \\
p_V&=&p_V-(A+B)\,d \cos{(b)}\cos{(l)},
\label{eq2}
\end{eqnarray}
where $d$ denotes the distance of a star 
and $l$ and $b$ are Galactic longitude and latitude, respectively. In this way
we have constructed proper motion velocities that reflect the true peculiar
velocities of the stars. For the Oort constants $A$ and $B$ we adopt the 
values determined by \cite{FeWh97}, $A = 14.82$ km/s/kpc and 
$B = - 12.37$ km/s/kpc.

\section{Determination of the Velocity Distribution Moments}\label{sec3}

\subsection{Projection Formalism}

The proper motion velocities, which we have introduced in the previous 
section, are formally defined as 
\begin{equation}
\left( \begin{array}{c} p_U \\ p_V \\ p_W \end{array} \right) =
\left( \begin{array}{c} U \\ V \\ W \end{array} \right) -
\left( (U,V,W),\vec{e}_r \right)\, \vec{e}_r ,
\label{eq3}
\end{equation}
where $\vec{e}_r$ is the unit vector pointing to the star under consideration, 
\begin{equation}
\vec{e}_r=\left( \begin{array}{c} \cos{(b)}\cos{(l)} \\ \cos{(b)}\sin{(l)} \\ 
\sin{(b)} \end{array} \right) .
\label{eq4}
\end{equation} 
Equation \ref{eq3} can be written as a set of algebraic equations 
\begin{equation}
\left( \begin{array}{c} p_U \\ p_V \\ p_W \end{array} \right) = {\mathcal{A}}
\cdot \left( \begin{array}{c} U \\ V \\ W \end{array} \right) .
\label{eq5}
\end{equation} 
The elements of the symmetric matrix ${\mathcal{A}}$ are given in Appendix A. 
The key assumption of the de-projection of the proper motion velocities is 
that the lines of sight towards the stars are statistically uncorrelated 
with the velocities of the stars \citep{DeBi98}. We thus take the 
statistical average of eq.~\ref{eq5} over the samples in each counting 
volume, as defined above $(\frac{1}{N}\sum_{n=1}^N)$
\begin{equation}
\Bigg \langle \begin{array}{c} p_U \\ p_V \\ p_W \end{array} \Bigg \rangle
= \langle {\mathcal{A}} \rangle
\cdot \Bigg \langle \begin{array}{c} U \\ V \\ W \end{array} \Bigg \rangle .
\label{eq6}
\end{equation} 
In contrast to eq.~\ref{eq5}, which cannot be 
inverted, eq.~\ref{eq6} can be inverted as
\begin{equation}
\Bigg \langle \begin{array}{c} U \\ V \\ W \end{array} \Bigg \rangle =
\langle {\mathcal{A}} \rangle^{-1}\cdot 
\Bigg \langle \begin{array}{c} p_U \\ p_V \\ p_W \end{array} \Bigg \rangle ,
\label{eq7}
\end{equation} 
which gives the three first moments of the velocity distribution. Since we
presume to have selected a sample from a relaxed population of stars in the 
Galactic disk, $\langle U \rangle$ and $\langle W \rangle$ should reflect simply
the solar motion
relative to the standard of the rest, $U_\odot=-\langle U \rangle$ and 
$W_\odot=-\langle W \rangle$. The quantity $\langle V \rangle$ is,
due to the asymmetry of
the $V$ velocity distribution, more negative than $-V_\odot$.

Next we consider the second moments of the velocity distribution. The 
components of the proper motion velocity of each star can be combined as 
\begin{eqnarray}
p_ip_j&=&\sum\limits_{k,m}{\mathcal{A}}_{ik}\upsilon_k
{\mathcal{A}}_{jm}\upsilon_m \\ &= & \frac{1}{2}
\sum\limits_{k,m}({\mathcal{A}}_{ik}{\mathcal{A}}_{jm}+
{\mathcal{A}}_{jk}{\mathcal{A}}_{im})\upsilon_k\upsilon_m \nonumber ,
\label{eq8}
\end{eqnarray}
where the indices stand for $i, j, k, m = U, V, W$ and $(\upsilon_U,
\upsilon_V, \upsilon_W) = (U, V, W)$. Again we take the ensemble average 
\begin{equation}
\langle p_ip_j \rangle =  \frac{1}{2}
\sum\limits_{k,m}\langle{\mathcal{A}}_{ik}{\mathcal{A}}_{jm}+
{\mathcal{A}}_{jk}{\mathcal{A}}_{im}\rangle  
\langle\upsilon_k\upsilon_m \rangle ,
\label{eq9}
\end{equation}
which is manifestly symmetric 
in the indices $i, j$. Formally, we set up a set of algebraic equations
\begin{equation}
\Bigg \langle \begin{array}{c} p_Up_U \\p_Up_V \\ p_Up_W \\ p_Vp_V \\
 p_Vp_W \\ p_Wp_W \end{array} \Bigg \rangle =
\langle {\mathcal{B}} \rangle\cdot 
\Bigg \langle \begin{array}{c} UU \\ UV \\ UW \\ VV \\ VW \\ WW
\end{array} \Bigg \rangle ,
\label{eq10}
\end{equation}
Since the lines of sight to the stars are statistically uncorrelated 
with their velocities, the matrix $\langle {\mathcal{B}}
\rangle$ is also not singular, and allows for each $Z$-slice the inversion
\begin{equation}
\Bigg \langle \begin{array}{c} UU \\ UV \\ UW \\ VV \\ VW \\ WW
\end{array} \Bigg \rangle=
\langle {\mathcal{B}} \rangle^{-1} \cdot 
\Bigg \langle \begin{array}{c} p_Up_U \\p_Up_V \\ p_Up_W \\ p_Vp_V \\
 p_Vp_W \\ p_Wp_W \end{array} \Bigg \rangle .
\label{eq11}
\end{equation}
The elements of matrix ${\mathcal{B}}$ are also given in Appendix A.

The velocity dispersions are given by 
$\sigma_{RR}^2=\langle UU \rangle - \langle
U \rangle^2,\,\sigma_{\phi\phi}^2=\langle VV \rangle - \langle
V \rangle^2,\,\sigma_{ZZ}^2=\langle WW \rangle - \langle
W \rangle^2 $ and we have the mixed moments 
$\sigma_{R \phi}^2=-(\langle UV \rangle - \langle U \rangle \langle
V \rangle), \sigma_{R Z}^2=-(\langle UW \rangle - \langle U \rangle \langle
W \rangle), \sigma_{\phi Z}^2=\langle VW \rangle - \langle V \rangle \langle
W \rangle$.
In the definition of the latter, we have switched from $U$ to $\dot{R}=-U$
which is, for instance, more convenient to use with the Jeans equations.

It seems intuitive that this inversion should work for an all-sky sample, but
less so in the case of the {\em SDSS} coverage. In order to test that the sky
coverage of the selected sample of stars is sufficient to reliably invert
eqns.~\ref{eq6} and \ref{eq10}, we have carried out a Monte-Carlo simulation. 
We have assigned to each star in the counting volumes a space velocity drawn 
randomly from the Schwarzschild distribution 
\begin{eqnarray}
f & \propto &
\exp{-\frac{1}{2} \left [\left(\frac{U-\langle U \rangle}{\sigma_{RR}}\right)^2 
 \right. } \\ & & \left. +
\left(\frac{V-\langle V \rangle}{\sigma_{\phi\phi}}\right)^2+ 
\left(\frac{W-\langle W \rangle}{\sigma_{ZZ}}\right)^2 \right] \nonumber
\label{eq12}
\end{eqnarray}
with $\langle U \rangle$ = -10 km/s, $\langle V \rangle$ = -26 km/s,
$\langle W \rangle$ = -7 km/s, $\sigma_{RR}$  = 45 km/s,  
$\sigma_{\phi\phi}$ = 32 km/s, and $\sigma_{ZZ}$ = 24 km/s. We have converted 
the space velocities to proper motion velocities and  
found that we could reconstruct, from just the samples contained in 
the {\em SDSS} footprint, both the assumed velocity offsets
and velocity dispersions to accuracies of 0.2 km/s, respectively.

\subsection{Results}

The de-projection procedure provides us with an observational estimate of the 
stellar velocity dispersion ellipsoid as a 
function of height above the Galactic midplane.
Results are given in Table 1 and illustrated in Fig.~\ref{fig2}. The error 
calculus for all entries in the table is described in Appendix B. As can be seen
in Fig.~\ref{fig2}, there is a mild scatter of the data points which we
ascribe to individual distance errors. 

\begin{deluxetable}{ccccccccccc}
\tablewidth{0pt}
\tablecaption{First and second moments of the velocity
              distribution as function of height above
              the Galactic midplane\label{tab1}}
\tablehead{\colhead{Z}   & \colhead{N} &  \colhead{$\langle U \rangle$} &
\colhead{$\langle V \rangle$} &
\colhead{$\langle W \rangle$} &
\colhead{$\sigma_{RR}$} &
\colhead{$\sigma_{\phi\phi}$} &
\colhead{$\sigma_{ZZ}$}  &
\colhead{${\langle UV \rangle}$} &
\colhead{$\langle UW \rangle$} &
\colhead{$\langle VW \rangle$} \\

$pc$ &  & $km/s$ & $km/s$ & $km/s$ &
 $km/s$ & $km/s$ & $km/s$ & ($km/s)^2$ & ($km/s)^2$ & ($km/s)^2$}
\startdata
 0--100  & 48\,075  & -8.62 & -20.04 & -7.10 & 32.4& 23.0  & 18.1  & 320.7 & 
 32.1 & 127.4 \\
  &   & 0.22 & 0.14 & 0.16 & 0.11 & 0.09 & 0.18 & 3.9 & 3.7  & 2.3\\[1mm]
 {\it CNS4}& {\it 600} & {\it -12.2} & {\it -22.3} & {\it -6.9} & {\it 38.2} &
  {\it 25.6}  & {\it 19.4}  & 151.1  & -26.7 &   -1.8 \\
&  & {\it 2.1} & {\it 1.5} &{\it 1.3} & {\it 1.7} & {\it 1.4} & {\it 1.3} &  -  & - & - \\[1mm]
{\it DB98} & $\sim$ {\it 1\,300} & {\it -10.44}& {\it -24.59} & {\it -8.17} & {\it 36.8} & {\it 26.7} & {\it 18.3}  & {\it 416.0} & {\it -86.1} & {\it 202.7}  \\
&  &  {\it 1.69} & {\it 1.71} & {\it 1.78} &{\it 1.18}  & {\it 1.72} & {\it 2.58} & -  &  - &  - \\[1mm]
100--200 & 280\,929 & -8.10 & -21.62 & -7.14 & 35.3 & 24.9 & 20.9 & 298.9 & 
7.1 & 143.3 \\
& & 0.09 & 0.06 & 0.08 & 0.04 &  0.04 &  0.08  & 1.7 & 1.7  &  1.2\\[1mm]
200--300 & 387\,749 & -8.05 & -22.37 & -7.49 & 38.5 & 27.2 & 24.9 & 292.9 &
-42.2 & 149.6 \\ 
& &  0.08 & 0.05 & 0.08 & 0.04 &  0.03 &  0.08 &  1.6 & 1.7  & 1.4\\[1mm]
300--400 & 360\,829  & -9.06 & -23.51 & -7.29 & 41.5 & 29.8 & 28.0 & 325.5 &
-75.9  & 137.0 \\
&  & 0.08 & 0.06 & 0.09 & 0.04 & 0.04 &  0.10 &  1.9 & 2.1  & 1.9\\[1mm]
400--500 & 302\,601  & -9.62 & -24.76 & -7.35 & 43.1 & 31.7 & 30.6 & 332.2 &
-125.4 & 135.6 \\
&  & 0.09 & 0.07 & 0.11 & 0.05 & 0.05 &  0.13 &  2.3 & 2.7  & 2.5\\[1mm]
500--600 & 237\,746  & -10.28 & -26.22 & -7.12 & 45.0 & 33.6 & 33.3 & 342.8 &
-181.1 & 122.3 \\
&  & 0.11 & 0.09 & 0.14 & 0.06 & 0.06 &  0.17 &  3.0 & 3.6  & 3.6\\[1mm]
600--700 & 168\,595  & -10.19 & -28.61 & -7.12 & 47.0 & 35.4 & 36.6 & 358.6 &
-232.5 & 132.3 \\
&  & 0.13 & 0.11 & 0.19 & 0.08 & 0.07 &  0.23 &  3.9 & 4.7  & 5.0\\[1mm]
700--800 & 112\,074  & -10.42 & -32.18 & -7.18 & 49.1 & 37.6 & 40.0 & 419.2 &
-263.0 & 166.8 \\
&  & 0.17 & 0.14 & 0.27 & 0.11 & 0.09 &  0.36 &  5.4 & 6.9  & 7.7\\
\enddata
\\
Col.~(1): distance interval, col.~(2): number of stars, cols.~(3) to (5): mean
velocity components and their errors (below), cols.~(6) to (8): velocity 
dispersions and their errors (below), cols.~(9) to (11): mixed moments of the
velocity ellipsoid and their errors (below). The errors represent only internal
errors. Systematic errors can be as large as 2-3 km/s.
\end{deluxetable}

In order to demonstrate the systematic effect of the adopted distance scale, we
also show in Fig.~\ref{fig2} kinematical results based on the faint 
normalization of the main sequence by \cite{Juri08}. As discussed in \S 2, this
distance scale is appropriate for metal-poor thick-disk stars. Since our sample
is presumed to be significantly contaminated by thick-disk stars
only in the most distant counting volumes, the alternative distance scale 
should be
actually used only in these counting volumes. As can be seen from the panels
on the right hand side of Fig.~\ref{fig2}, switching the distance scale leads to
differences of the the mean motions and velocity dispersions of 1 to 2 km/s,
with the exception of $\sigma_{RR}$, which deviates by up to 3 km/s. These
changes are unexpectedly small, but can be understood from closer inspection of
Fig.~\ref{fig-1}. The lowest distance bin is populated by comparatively red
stars with typical colors $1.1 \lesssim (r-i)_0 \lesssim{<}1.5$. As can be seen
from Fig.~\ref{fig-1}, the difference $M_r({\rm faint})-M_r({\rm bright})$
ranges in this color range from $0.2$ to -$0.2$, mag so that in the statistical
average no change of the kinematic results is expected. In the most distant bin,
$0.5 < (r-i)_0 < 1$, the difference $M_r({\rm faint})-M_r({\rm bright})$ is
about 0.2 mag, which corresponds to a distance scale contraction of 9\%.
However, as a result of the reduced distance estimates, stars will wander into
the $Z$-slice from above, and some stars will drop through the bottom of the
counting volume. As can be seen from Fig.~\ref{fig0b}, the incoming stars (from
above) will be systematically bluer than the stars which they replace (from
below). The former will also be systematically brighter, which compensates to a
considerable degree the switch from the bright to the faint calibration of the
absolute magnitudes. Thus, the systematic errors of the kinematical data
introduced by systematic errors in the distance scale are only of order $\sim$ 
1 to 2 km/s.

As an independent consistency check on our results, we have retrieved from the
SpecPhoto database of {\em SDSS DR-7} the radial velocities of all stars at
galactic latitudes $b \geq 70^\circ$ in the color window $1.0 \leq g-r \leq
1.5$ and $0.3 \leq r-i \leq 1.6$. Since these stars lie close to the Galactic
pole, their radial velocities reflect their $W$ velocity components. The
resulting mean velocities and velocity dispersions are summarized as a function
of
height above the Galactic midplane in Table 2. As can be seen from a comparison
with Table 1, the vertical velocity dispersions determined in this way are 
in excellent agreement with those estimated from proper motions alone.

\begin{deluxetable}{cccccc}
\tablewidth{0pt}
\tablecaption{First and second moments of the distribution of 
spectroscopically measured line-of-sight velocities$^\dag$ of M dwarfs 
at latitudes $b \geq 70^\circ$ \label{taba0}}
\tablehead{\colhead{Z}   & \colhead{N} &  \colhead{$\langle V_{los} \rangle$} & 
\colhead{$\epsilon_{\langle V_{los} \rangle}$} &  
\colhead{$\sigma_{V_{los}}$} & 
\colhead{$\epsilon_{\sigma_{V_{los}}}$} \\
 pc     &     & km/s & km/s & km/s & km/s}
\startdata
 0--200  & 272 & -9.5  & 0.6 & 24.0 & 1.0\\
200--400 & 770 & -12.0 & 0.4 & 27.6 & 0.7\\
400--600 & 247 & -8.5  & 0.5 & 29.7 & 1.3\\
600--800 &  64 & -16.0 & 2.0 & 39.6 & 3.5\\
\enddata
\\
$^\dag$SpecPhoto data from {\em SDSS DR-7}
\end{deluxetable}

The second row in Table 1 lists, for comparison, kinematical data for late-type
stars\footnote{Groups 2-5 as defined in \cite{JaWi97}} reproduced from
the {\em CNS4} (Jahrei{\ss}, in preparation), which contains data for 
stars within a distance of 25 pc from the Sun. With the exception of the mean 
radial velocity $\langle U \rangle$, the other first and second moments of the 
velocity distribution determined in the Z = 0-100 pc sample of {\em SDSS} 
stars are remarkably consistent with the relatively tiny sample of nearby 
stars. The third row in Table 1 gives the results of \cite{DeBi98} in their
reddest bin, which corresponds to K-type stars. Again, there is good agreement
between 
their results and the results in our low latitude sample. We note as well
that \cite{Velt08} find, from their analysis of the {\em RAVE} survey 
\citep{Stei06} $\langle U \rangle$ = -8.5 km/s, which is very close to our
{\rm SDSS} result. Moreover, we can compare our results with the observations of
the SLoMaSS stars by \cite{Boc07}. For this purpose we have combined all 
$1\,380\,183$
stars with $0 < Z < 500$ pc and derived with the statistical de-projection
method velocity dispersions ($\sigma_{RR}, \sigma_{\phi\phi},\sigma_{ZZ}$) = 
(40.0, 28.5, 25.4) km/s. These are in excellent agreement with the measurements
by \cite{Boc07} (39.0, 30.0, 24.8) km/s.

The distribution of the $U$-velocity components of the stars is expected to be
symmetrical with respect to the local standard of rest (LSR). Thus the mean 
velocity $\langle U \rangle$ reflects the solar motion relative to the LSR,
$\langle U \rangle = - U_\odot$. But as can be seen from Table 1, 
we find a slight, but statistically significant, 
decrease of $\langle U \rangle$ with increasing height above the midplane. 
This might indicate some radially inward streaming of the local stars relative
to the rest of the disk, perhaps induced by a spiral arm. 

The mean rotation velocity $\langle V \rangle$ does not simply reflect the solar
motion $-V_\odot$, but is much more negative due to the asymmetrical shape of
the distribution of the $V$-velocity components of the stars. The solar motion 
in $V$ alone would be reflected by a subsample of young stars with very small
velocity dispersions (cf.~\citealt{DeBi98}), which we cannot isolate in our
data, because we do not know the individual space velocities of the stars. 
The fourth column of Table 1 shows that the mean rotation velocity of stars 
slows down systematically with height above the midplane, as 
$\Delta \langle V \rangle / \Delta\sigma_{RR}^2$ = -0.0089 (km/s)$^{-1}$, 
and ~$\Delta \langle V \rangle / \Delta Z$ = - 0.017 km/s/kpc.  
The latter is expected due to the asymmetric drift effect \citep{BiTr87}.
A quantitative interpretation of this effect is not possible on the basis of 
the present data. For Jeans modeling of the Milky Way disk, the radial 
gradient of the $\sigma_{RR}$ velocity dispersion must be known. 
Unfortunately, this cannot be measured using stars from the {\em SDSS DR-7}
sample, but we hope to address this problem in the near future.

The quantity $\langle W \rangle = - W_\odot$ stays constant with 
height above the midplane.

The velocity dispersions $\sigma_{RR}$,  $\sigma_{\phi \phi}$, 
and $\sigma_{ZZ}$ rise systematically with height above the midplane, $Z$. 
This rise is expected due to two effects. Even the relaxed population of M-type
stars represents an age mix of young and old stars. Since the work of
\cite{SpSc51}, it is well known that older stars exhibit, as a result
of `disk heating', significantly larger velocity dispersions than younger stars
(see ~\citealt{Fuch01} for a review). Stars with larger vertical $\sigma_{ZZ}$
velocity dispersions can ascend higher into the Galactic gravitational
potential, so that one expects to find
stars with higher velocity dispersions at greater heights above the midplane 
than closer to the midplane (\citealt{FuWi87}, \citealt{Wes08}). Furthermore,
our sample transitions
with increasing height above the midplane from the old thin-disk to the thick-
disk population. Thick-disk stars have velocity dispersions on the order 
of $(\sigma_{RR},\sigma_{\phi\phi}, \sigma_{ZZ}) $ = (45-65, 40-50, 30-40) km/s
(\citealt{ChBe00}, \citealt{AlCu05},
\citealt{Hol07}), which are only slightly larger than what we actually measure.
We have estimated, from the 50\% contour level of the velocity distribution of
red clump giants at altitudes between 500 pc and 1500 pc above the midplane
derived by \cite{Sie08} with {\em RAVE DR-2} data, velocity dispersions of about
$(\sigma_{RR}, \sigma_{\phi\phi}, \sigma_{ZZ})$ $\approx$ (79, 56, 35) km/s.
The vertical velocity dispersion is consistent with our measurement in the 
highest altitude bin, whereas the planar velocity dispersions are significantly 
higher than our measurements, and, particularly in the $U$ velocity component,
even higher than what is commonly ascribed to thick-disk stars. We note, 
however, that such high $\sigma_{RR}$ and $\sigma_{\phi\phi}$ dispersions are
comparable
with the velocity dispersions ascribed by \cite{Boc07} to the high-dispersion
wings of the velocity distribution of the stars in their sample. Thus, the 
sample of \cite{Sie08}, with a `ceiling' at $Z$ = 1500 pc, seems to be 
contaminated by hot metal-weak thick disk \citep{ChBe00} and halo stars;
the tilt of the velocity ellipsoid which they measure might not 
reflect that of disk stars.

The rotational lag of the stars is tied by the asymmetric drift effect to the 
radial velocity dispersion $\sigma_{RR}$, so that $-\langle V \rangle$ 
increases with $Z$, precisely as observed. As explained above, a quantitative 
modeling of this effect is not possible on the basis of the present data alone. 

Owing to the richness of our data, we were able to split them up into subsets
based on color. As a test, we have selected only those stars with colors in the
range 0.49 $\leq$ $(r-i)_0$ $\leq$ 0.95, where our sample is volume complete
(cf.~Fig.~\ref{fig0b}). If we use only these stars in the de-projection of the
proper motion velocities, we find the first and second moments of the velocity
distribution illustrated in Fig.~\ref{figxx}. A comparison with Fig.~\ref{fig2}
shows that the kinematical results inferred from the `blue' subset are entirely
consistent with that of the total data set. Only in the $\langle V \rangle$
velocity component do we find, close to the midplane, an offset of 4 km/s. 

Next we discuss the off-diagonal elements of the velocity dispersion tensor.
The mixed moment $\sigma_{R\phi}$ is illustrated in Fig.~\ref{fig4} as a 
function of height above the Galactic midplane. The inset of Fig.~\ref{fig4} 
shows the implied vertex deviation
\begin{equation}
\psi = -\frac{1}{2} \arctan{\frac{2\, \sigma_{R\phi}^2}
{\sigma_{RR}^2-\sigma_{\phi\phi}^2}}\,.
\label{eq14}
\end{equation}
As can be seen from Fig.~\ref{fig4}, the vertex deviation is non-negligible,
even
if it falls slightly with height above the midplane. Close to the midplane our
measurement is consistent with the result of \cite{DeBi98}. Since our sample of
stars is supposed to be drawn from a dynamically relaxed population of the Milky
Way, the vertex deviation is almost certainly due to dynamical effects induced
by non-axisymmetric structures in the Galactic disk.
  
The moment $\sigma_{RZ}^2$ can be used to determine the apparent tilting angle 
of the velocity ellipsoid in the meridional plane spanned by $U$ and $W$,
\begin{equation}
\alpha = -\frac{1}{2} \arctan{\frac{2\, \sigma_{RZ}^2}
{\sigma_{RR}^2-\sigma_{ZZ}^2}}\,.
\label{eq13}
\end{equation}
The mixed moment itself is shown as $-\sqrt{\sigma_{RZ}^2}$ in Fig.~\ref{fig3}
as a function of height above the Galactic midplane. The resulting tilting angle
$\alpha$ is depicted in the inset of Fig.~\ref{fig3}. Close to the midplane we 
find $\alpha \approx 0$ implying that the velocity ellipsoid points towards the
Galactic Center. Above the midplane the velocity ellipsoid is tilted towards 
the Galactic center, however, by a much larger angle than previously thought
(\citealt{BiTr87}, \citealt{Gil89}, \citealt{Ken91}). 
At a height of $Z$ = 750 pc the 
velocity ellipsoid is tilted by $\alpha$ = $-20^\circ$, whereas one would
expect $\alpha$ = $-5.4^\circ$ if the velocity ellipsoid was pointing straight
towards the Galactic center. In order to assess the plausibility of this 
unexpected result we address each of the three input parameters of
eq.~\ref{eq13} in turn. We have already tested above our measurements of the
vertical velocity dispersions $\sigma_{ZZ}$ using a completely different sample
of M stars, for which the line-of-sight velocities in the direction towards the
Galactic pole are known. The radial velocity dispersions $\sigma_{RR}$ can be
checked in a similar way. We have determined the first and second moments of the
distribution of the $p_U$  and $p_V$ proper motion velocity components of those
stars in our sample which lie in the cone $b \geq 70^\circ$. In the upper most
counting volume, at 700 pc $\leq Z \leq$ 800 pc, we find 
$\sqrt{ \langle p_U p_U \rangle - \langle p_U \rangle^2}$ = 47.9$ \pm$ 0.2 km/s
and $\sqrt{ \langle p_V p_V \rangle - \langle p_V \rangle^2}$ = 
36.1 $\pm$ 0.2 km/s, which are in excellent agreement with the result of our 
de-projection formalism (see ~Table 1). There is no independent test of the 
mixed moment $\sigma_{RZ}^2$ within our data at this altitude, but we
note that the tilting angle of the velocity ellipsoid of $\alpha = 7.3^\circ$
reported by \cite{Sie08} implies $\sigma_{RZ}^2 \approx 650$ (km/s)$^2$,
which is even more than what we find. At mid altitudes, around 400 pc
above the midplane, the sample of M stars with known radial velocities
(described above) provides an independent check on the mixed moment  
$\sigma_{RZ}^2$. We have drawn from the SpecPhoto database of {\em SDSS DR-7} a 
set of 201 stars with $b \geq 70^\circ$ at heights of 400 $<$ $Z$ $<$ 600 pc
above the midplane, with a mean of $\langle Z \rangle$ = 450 pc. Since the stars
lie in the cone $b \geq 70^\circ$, the line-of-sight velocities reflect their
$W$ velocity components; we obtain $\sigma_{V_{los}}$ = 29.2 $\pm$ 0.7 km/s.
From the proper motion velocities $p_U$ we derive
$\sqrt{ \langle p_U p_U \rangle - \langle p_U \rangle^2}$ = 46.8$ \pm$ 1.8 km/s.
We have then cross-correlated the $V_{los}$ and $p_U$ velocity components, 
giving $\langle V_{los} p_U \rangle$ = -275.5 $\pm$ 46 (km/s)$^2$, which
implies a
tilting angle of $\alpha $ = 11.2 $\pm$ 1.7 degrees. If the velocity ellipsoid
were pointing towards the Galactic center, one would expect $\alpha$ = 3.2
degrees. A comparison of these independent kinematic data with the seventh and
eighth row of Table 1 and Fig.~\ref{fig3} shows excellent agreement with the
results from the statistical de-projection of the proper motion velocities.
We conclude from this discussion that our measurement of the tilting angle of
the velocity dispersion tensor should be quite robust.
 
Finally, we determine the epicyclic ratio, which is related to the slope of the
Galactic rotation curve as
\begin{equation}
\frac{d{\rm ln}v_c}{d{\rm ln} r}= 2\,\frac{\sigma_{\phi\phi}^2}{\sigma_{RR}^2}
- 1\,,
\label{eq15}
\end{equation}
Close to the midplane, we find a value of $ \frac{d{\rm ln}v_c}{d{\rm ln} r}$ =
-0.006 $\pm$ 0.016. which indicates a flat Galactic rotation curve in the solar
neighborhood. Note, however, that the quoted error reflects only internal
errors. Equation (\ref{eq15}) neglects the asymmetric shape of the
distribution of the $V$ velocity components of the stars. In order to correct
for this simplification, third moments of the velocity distribution must be 
known (\citealt{CuBi94}), which we have not measured here.




\section {Conclusions}\label{sec4}

We have analyzed the velocity distribution of stars in the Galactic old 
thin-disk population using stellar data from the {\em Seventh Data Release of
the Sloan Digital Sky Survey}. We have constructed a sample of about 
2 million late-type stars in eight cylindrical counting volumes, each of which 
has a height of 100 pc and a diameter of 2 kpc, respectively, and 
which were stacked onto one another. For each star we have determined a 
photometric distance estimate, and combined this with their available proper
motions to obtain tangential velocities. These were then treated as pseudo 
space velocities. We have demonstrated with Monte Carlo simulations that 
our sample provides sufficient lines of sight through the velocity distribution
of the stars that the pseudo space velocities allow for a precise statistical
determination of the first and second moments of the three dimensional velocity
distribution. Close to the Galactic midplane we find $\langle U \rangle =$ -8.6
$\pm$ 0.2 km/s and $\langle W \rangle =$ -7.1 $\pm$ 0.2 km/s, which directly
reflect the motion of the sun relative to the Local Standard of Rest. 
The mean $V$ velocity component, $\langle V\rangle =$ -20.0 $\pm$ 0.1 km/s, is
much more negative than just the reflex of the solar motion, due to
the asymmetric shape of the $V$ velocity distribution. Close to the midplane, we
obtain estimates of the velocity dispersions $\sigma_{RR}=$ 32.4 $\pm$ 0.1 km/s,
$\sigma_{\phi \phi}=$ 23.0 $\pm$ 0.1 km/s, and $\sigma_{ZZ}=$ 18.1 $\pm$ 0.2
km/s, respectively. With the exception of $\langle U \rangle$, all of these
results are consistent with other determinations, such as analyses of the {\em
Catalogue of Nearby Stars} or {\em Hipparcos} stars by \cite{DeBi98}. However,
due to the large number of stars we reach an unprecedented precision, up to ten
times better than previously achieved. Moreover, we have a sufficient sample to
trace all quantities as a function of height above the Galactic midplane. The
velocity dispersions exhibit a distinct increase with height above the midplane.
This is a clear indication that our sample of M stars contains a mixture of
young stars with low velocity dispersions and old stars with high velocity
dispersions, because only the latter ascend high above the midplane. Since the
rotational lag of the stars is tied by the asymmetric drift effect to the
radial velocity dispersion, -$\langle V \rangle$ must increase also with height
above the midplane, precisely as observed.

We have then studied the orientation of the velocity ellipsoid in the
meridional planes spanned by the $U$, $V$, and $W$ velocity components.
We find a vertex deviation of the velocity ellipsoid of about 20$^\circ$ to
25$^\circ$. Close to the Galactic midplane the ellipsoid is oriented parallel to
the  Galactic plane. At a height of 750 pc above the midplane the velocity 
ellipsoid is tilted by an angle of about $\alpha = -20^\circ$. This is more 
than thought previously, because a tilting angle of $\alpha  = -5.4^\circ $ is
expected at that altitude, if the velocity ellipsoid points straight at the
Galactic Center.

Finally, we determined the epicyclic ratio which indicates, close to the
midplane, a flat shape of the Galactic rotation curve in the solar neighborhood.
 
We have refrained in this paper from any detailed dynamical modelling of our
data but hope to address this topic in the near future.

\acknowledgements

Funding for the SDSS and SDSS-II has been provided by the Alfred P.~Sloan 
Foundation, the Participating Institutions, the National Science Foundation, 
the U.S.~Department of Energy, the National Aeronautics and Space 
Administration, the Japanese Monbukagakusho, the Max Planck Society, and the 
Higher Education Funding Council for England. The SDSS Web Site is
http://www.sdss.org/.

The SDSS is managed by the Astrophysical Research Consortium for the 
Participating Institutions. The Participating Institutions are the 
American Museum of Natural History, Astrophysical Institute Potsdam, 
University of Basel, University of Cambridge, Case Western Reserve University,
University of Chicago, Drexel University, Fermilab, the Institute for Advanced
Study, the Japan Participation Group, Johns Hopkins University, the Joint 
Institute for Nuclear Astrophysics, the Kavli Institute for Particle 
Astrophysics and Cosmology, the Korean Scientist Group, the Chinese Academy of
Sciences (LAMOST), Los Alamos National Laboratory, the Max-Planck-Institute for 
Astronomy (MPIA), the Max-Planck-Institute for Astrophysics (MPA), New Mexico 
State University, Ohio State University, University of Pittsburgh, 
University of Portsmouth, Princeton University, the United States Naval 
Observatory, and the University of Washington.

\section*{Appendix A: The Matrices ${\mathcal{A}}$ and ${\mathcal{B}}$} 
\label{sec5}

The elements of matrix ${\mathcal{A}}$ are given by
\begin{eqnarray}
{\mathcal{A}}_{UU}&=&1-{\rm cos}^2(b)\,{\rm cos}^2(l) \nonumber \\
{\mathcal{A}}_{UV}&=&-{\rm cos}^2(b)\,{\rm cos}(l)\,{\rm sin}(l)=
{\mathcal{A}}_{VU}
\nonumber \\
{\mathcal{A}}_{UW}&=&-{\rm sin}(b)\,{\rm cos}(b)\,{\rm cos}(l)
={\mathcal{A}}_{WU}
\nonumber \\
{\mathcal{A}}_{VV}&=&1-{\rm cos}^2(b)\,{\rm sin}^2(l) \nonumber \\
{\mathcal{A}}_{VW}&=&-{\rm sin}(b)\,{\rm cos}(b)\,{\rm sin}(l)
={\mathcal{A}}_{WV}
\nonumber \\
{\mathcal{A}}_{WW}&=&{\rm cos}^2(b) \,.
\label{eqA1}
\end{eqnarray}
The elements of matrix  ${\mathcal{B}}$ are defined straightforwardly by 
eq.~\ref{eq9}, but are a bit cumbersome to write out explicitely. We consider
${\mathcal{B}}(nr,nc)$, where $nr$ and $nc$ are the numbers of the rows and 
columns in the matrix ${\mathcal{B}}$. Next we introduce two tables of indices 
$icol$ and $irow$ and
write, if $nc$ equal 1, 4, or 6
\begin{eqnarray}
{\mathcal{B}}(nr,nc)&=&
\frac{1}{2}\left[ {\mathcal{A}}(irow(nr,1),icol(nc,1)) \right. \nonumber \\
 & & *{\mathcal{A}}(irow(nr,2),icol(nc,2)) \\
 & & +{\mathcal{A}}(irow(nr,3),icol(nc,3))\nonumber \\
 & & \left. *{\mathcal{A}}(irow(nr,4),icol(nc,4)) \right] \nonumber \,. 
\label{eqA2}
\end{eqnarray}
For the other combinations we have
\begin{eqnarray}
{\mathcal{B}}(nr,nc)&=&
\frac{1}{2}\left[ {\mathcal{A}}(irow(nr,1),icol(nc,1)) \right. \nonumber \\
 & & *{\mathcal{A}}(irow(nr,2),icol(nc,2)) \nonumber \\
 & & +{\mathcal{A}}(irow(nr,3),icol(nc,3))\nonumber \\
 & & *{\mathcal{A}}(irow(nr,4),icol(nc,4))  \\ 
 & & +{\mathcal{A}}(irow(nr,1),icol(nc,5))\nonumber \\
 & & *{\mathcal{A}}(irow(nr,2),icol(nc,6))  \nonumber \\ 
 & & +{\mathcal{A}}(irow(nr,3),icol(nc,7))\nonumber \\
 & & \left. *{\mathcal{A}}(irow(nr,4),icol(nc,8)) \right]  \nonumber  \,.
\label{eqA3}
\end{eqnarray}
\begin{deluxetable}{lcccc}
\tablewidth{0pt}
\tablecaption{Index table irow(nr,j) \label{taba1}}
\tablehead{\colhead{nr}   & \colhead{j=1} &  \colhead{j=2} & 
\colhead{j=3} &  \colhead{j=4}}
\startdata
1 & 1 & 1 & 1 & 1\\
2 & 1 & 2 & 2 & 1\\
3 & 1 & 3 & 3 & 1\\
4 & 2 & 2 & 2 & 2\\
5 & 2 & 3 & 3 & 2\\
6 & 3 & 3 & 3 & 3\\
\enddata
\end{deluxetable}
\begin{deluxetable}{lcccccccc}
\tablewidth{0pt}
\tablecaption{Index table icol(nc,j) \label{taba2}}
\tablehead{\colhead{nc}   & \colhead{j=1} &  \colhead{j=2} & 
\colhead{j=3} &  \colhead{j=4} & \colhead{j=5} & \colhead{j=6} &
\colhead{j=7}  & \colhead{j=8}}
\startdata
1 & 1 & 1 & 1 & 1 & - & - & - & -\\
2 & 1 & 2 & 1 & 2 & 2 & 1 & 2 & 1\\
3 & 1 & 3 & 1 & 3 & 3 & 1 & 3 & 1\\
4 & 2 & 2 & 2 & 2 & - & - & - & -\\
5 & 2 & 3 & 2 & 3 & 3 & 2 & 3 & 2\\
6 & 3 & 3 & 3 & 3 & - & - & - & -\\
\enddata
\end{deluxetable}
The indices $irow$ and $icol$ of matrix ${\mathcal{A}}$ stand for 
$irow, icol = 1, 2, 3, = U, V, W$, respectively.


\section*{Appendix B: Error Calculus} \label{sec6}

We consider first the moments  $\langle U \rangle$, $\langle V \rangle$,
and  $\langle W \rangle$. Each proper motion velocity component has an
individual error $\epsilon_U$,  $\epsilon_V$,  $\epsilon_W$. The error
of the statistically averaged proper motion velocity components is given by
\begin{equation}
\epsilon^2_{\langle p_U \rangle, \langle p_V \rangle,\langle p_W \rangle}
=\frac{1}{N^2}\sum _{n=1}^N \epsilon^2_{p_{U,n},p_{V,n},p_{W,n}} \,,
\label{eqB1}
\end{equation}
where $N$ denotes the number of stars in the counting volume.
We neglect positional errors so that matrix elements of 
$\langle {\mathcal{A}} \rangle$ have no errors. The errors of the first moments
of the velocity distribution follow then as
\begin{eqnarray}
\epsilon^2_{\langle U \rangle}=(\langle {\mathcal{A}_{UU}}\rangle^{-1})^2
\epsilon^2_{\langle p_U \rangle}&+&(\langle {\mathcal{A}_{UV}}\rangle^{-1})^2
\epsilon^2_{\langle p_V \rangle} \nonumber \\ & &
+(\langle {\mathcal{A}_{UW}}\rangle^{-1})^2
\epsilon^2_{\langle p_W \rangle} \nonumber \\
\epsilon^2_{\langle V \rangle}=(\langle {\mathcal{A}_{VU}}^{-1}\rangle)^2
\epsilon^2_{\langle p_U \rangle}&+&(\langle {\mathcal{A}_{VV}}^{-1}\rangle)^2
\epsilon^2_{\langle p_V \rangle} \\ & &
+(\langle {\mathcal{A}_{VW}}\rangle^{-1})^2
\epsilon^2_{\langle p_W \rangle} \nonumber \\
\epsilon^2_{\langle W \rangle}=(\langle {\mathcal{A}_{WU}}^{-1}\rangle)^2
\epsilon^2_{\langle p_U \rangle}&+&(\langle {\mathcal{A}_{WV}}^{-1}\rangle)^2
\epsilon^2_{\langle p_V \rangle} \nonumber \\ & &
+(\langle {\mathcal{A}_{WW}}\rangle^{-1})^2
\epsilon^2_{\langle p_W \rangle} \nonumber \,.
\label{eqB2}
\end{eqnarray}
The errors of the statistical averages of the combinations of the proper
motion velocity components are given by
\begin{eqnarray}
\epsilon^2_{\langle p_U p_U \rangle}& = &\frac{4}{N^2} \sum_{n=1}^N
p^2_{U,n} \epsilon^2_{p_{U,n}} \nonumber \\
\epsilon^2_{\langle p_U p_V \rangle}& = &\frac{1}{N^2} \sum_{n=1}^N
p^2_{V,n} \epsilon^2_{p_{U,n}}+ p^2_{U,n} \epsilon^2_{p_{V,n}}\nonumber\\
\epsilon^2_{\langle p_U p_W \rangle} &=& \frac{1}{N^2} \sum_{n=1}^N
p^2_{W,n} \epsilon^2_{p_{U,n}}+ p^2_{U,n} \epsilon^2_{p_{W,n}}\nonumber\\
\epsilon^2_{\langle p_V p_V \rangle} &=& \frac{4}{N^2} \sum_{n=1}^N
p^2_{V,n} \epsilon^2_{p_{V,n}} \\
\epsilon^2_{\langle p_V p_W \rangle} &=& \frac{1}{N^2} \sum_{n=1}^N
p^2_{W,n} \epsilon^2_{p_{V,n}}+ p^2_{V,n} \epsilon^2_{p_{W,n}}\nonumber\\
\epsilon^2_{\langle p_W p_W \rangle} &=& \frac{4}{N^2} \sum_{n=1}^N
p^2_{W,n} \epsilon^2_{p_{W,n}} \nonumber 
\label{eqB3}
\end{eqnarray}
The errors of the second moments of the velocity distribution
follow then as
\begin{eqnarray}
\epsilon^2_{\langle UU \rangle} &=& (\langle {\mathcal{B}_{11}}\rangle^{-1})^2
\epsilon^2_{\langle p_U p_U \rangle}+(\langle {\mathcal{B}_{12}}\rangle^{-1})^2
\epsilon^2_{\langle p_U p_V \rangle} \nonumber \\ 
&+&(\langle {\mathcal{B}_{13}}\rangle^{-1})^2 \epsilon^2_{\langle p_U p_W 
\rangle}+(\langle {\mathcal{B}_{14}}\rangle^{-1})^2 
\epsilon^2_{\langle p_V p_V \rangle} \\ 
&+&(\langle {\mathcal{B}_{15}}\rangle^{-1})^2 
\epsilon^2_{\langle p_V p_W \rangle}
+(\langle {\mathcal{B}_{16}}\rangle^{-1})^2 
\epsilon^2_{\langle p_W p_W \rangle}
\nonumber \\ 
\epsilon^2_{\langle UV \rangle} &=& (\langle {\mathcal{B}_{21}}\rangle^{-1})^2
\epsilon^2_{\langle p_U p_U \rangle}+(\langle {\mathcal{B}_{22}}\rangle^{-1})^2
\epsilon^2_{\langle p_U p_V \rangle}+ ... \nonumber
\end{eqnarray}
and so on.

\clearpage



\begin{figure*}
\includegraphics[angle=0,scale=1.0]{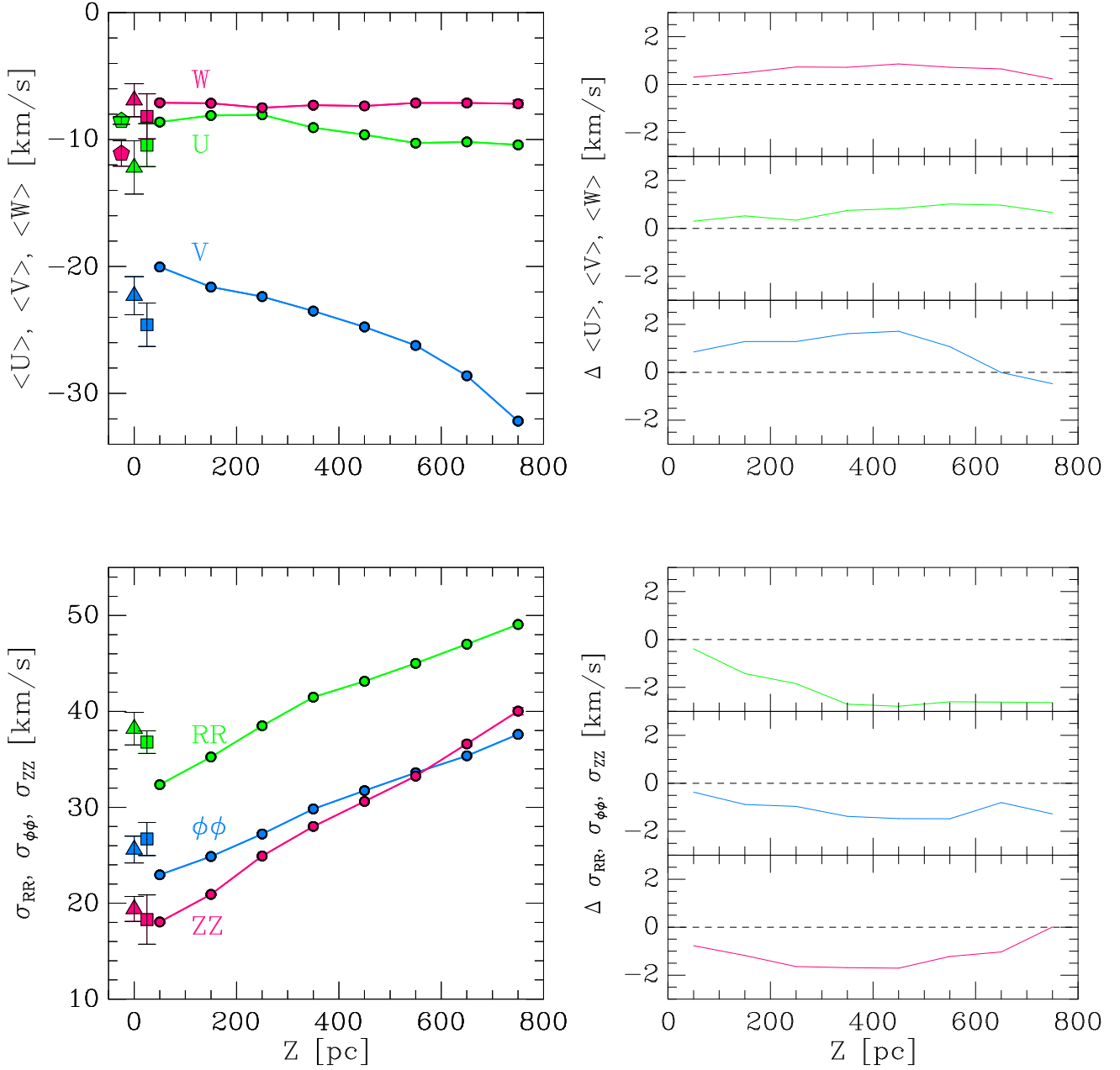}
\caption{Top left panel: The first moments $\langle U \rangle $, 
$\langle V \rangle$, and
$\langle W \rangle $ as a function of height $Z$ above the Galactic midplane
(solid dots), as determined in this work. Solar neighborhood data (triangles)
and the results of \cite{DeBi98} (squares) and  \cite{Velt08} (pentagons)
are indicated at $Z$ = 0.
Bottom left panel: The velocity dispersions $\sigma_{RR}$, 
$\sigma_{\phi \phi}$, and $\sigma_{ZZ}$ as a function of height 
$Z$ above the Galactic midplane (solid dots) as determined in this work. 
The solar neighborhood data (triangles) and results of \cite{DeBi98} 
(squares) are indicated at $Z$ = 0.
Top right panel: Differences of the mean velocities $\langle U \rangle $, 
$\langle V \rangle$, and $\langle W \rangle $ (color coded as in the left
panel) determined with the faint 
normalization of the absolute magnitudes minus the mean velocities determined
with the bright normalization \citep{Juri08}.
Bottom right panel: Same as upper panel, but for the velocity dispersions
$\sigma_{RR}$, $\sigma_{\phi \phi}$, and $\sigma_{ZZ}$. The right panels
indicate that the systematic errors of our results are of the order of 2 to 3
km/s.}
\label{fig2}
\end{figure*}

\begin{figure}
\plotone{figur45.ps}
\caption{Same as Fig.~\ref{fig2}, but using only stars with colors $0.49 \leq
(r-i)_0 \leq 0.9$. The systematic errors of our results are of the order of 2 to
3 km/s.}
\label{figxx}
\end{figure}

\begin{figure}
\plotone{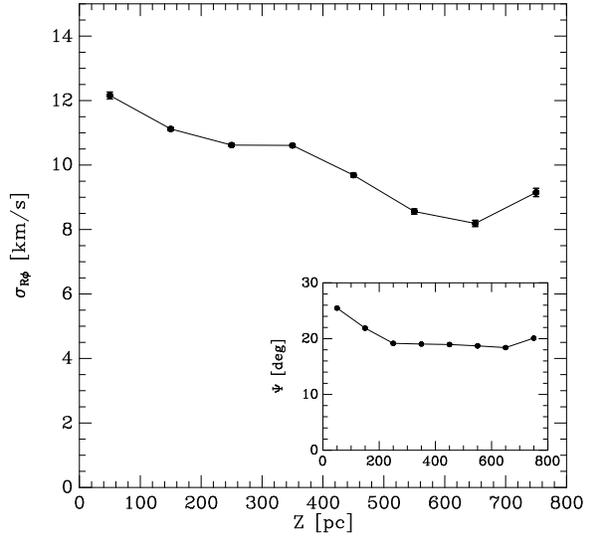}
\caption{The mixed element $-\sigma_{R\phi}$ of the velocity dispersion 
         tensor as a function of height above the Galactic midplane $Z$.
         The inset shows the vertex deviation of the velocity ellipsoid.}
\label{fig4}
\end{figure}

\begin{figure}
\plotone{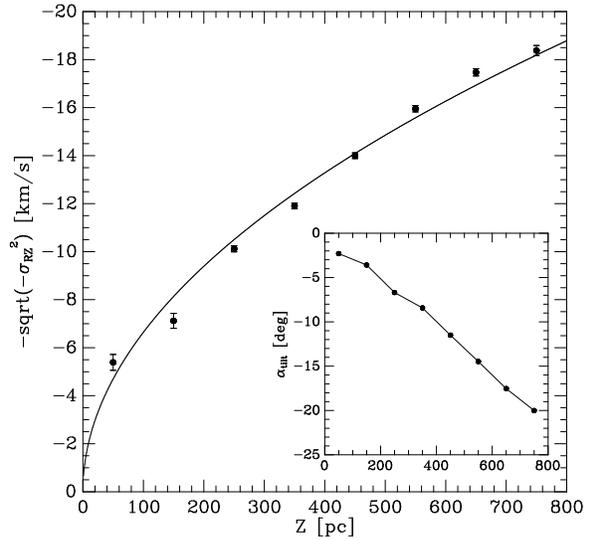}
\caption{The mixed element $-\sqrt{\sigma_{RZ}^2}$ of the velocity dispersion 
         tensor as a function of height above the Galactic midplane $Z$.
         The solid line line is a polynomial fit. The inset shows the
         apparent tilting angle of the velocity ellipsoid.}
\label{fig3}
\end{figure}

\end{document}